\documentclass[onecolumn,pra]{revtex4}
\usepackage{bm}
\usepackage{epsfig}
\usepackage{amsmath}
\begin{document}
\title{Barrierless electronic relaxation in solution - two state model with exact analytical solution in time domain}
\author{Aniruddha Chakraborty \\
School of Basic Sciences, Indian Institute of Technology Mandi,\\
Kamand, Himachal Pradesh, 175075, India}
\date{\today }
\begin{abstract}
\noindent  We propose an analytical method for solving the problem of electronic relaxation in solution in time domain, modelled by a particle undergoing diffusion under the influence of two coupled potentials. The coupling between the two potentials is assumed to be represented by a Dirac delta function of arbitrary position and strength. Smoluchowskii equation is used model the diffusion motion on both the potentials. We report an analytical expression for survival probability in time domain. This is the first time analytical solution in time domain is derived and this method can be used to solve problems involving other potentials.     
\end{abstract}
\maketitle
\noindent The problem of electronic relaxation of a molecule immersed in a polar solvent is studied both experimentally and theoretically \cite{Rice-Book,Hynes-Rev,Wolynes,Sadhukhan,Sutin,Sumi,Fleming,Oxtoby,Bagchi,Szabo,Kls,Skg,Hynes}. In laboratory, using a light of appropriate frequency, a molecule in polar solvent is electronically excited. This electronically excited molecule can relax back to the ground electronic state either by radiative mechanism or by no0n-radiative mechanism. Before electronic excitation the molecule is in the ground electronic state and it's molecular configuration changes randomly as a result of interaction between the molecule and the polar solvent. After interaction with light, the molecule is in the elctronically excited state, again in this state as well, molecular configuration changes randomly as a result of interaction between the excited molecule and the polar solvent. As the molecular configuration changes, from some specific configuration the molecule may undergo non-radiative decay to the ground electronic state. The molecule undergoes radiative decay from all molecular configurations with equal rate. From theoretical point of view, the problem is to calculate the probability that the molecule will still be in the excited state after time $t$. Now each molecular configuration can be modelled by position of a point particle, therefore random changes in molecular configuration can be modelled by 
random motion of a point particle on the relevant potential. In the following we use one dimensional Smoluchowski equation to model the random motion of the point particle, where the relevant coordinate is represented by $x$. In the following we use $P_{e}(x,t)$ to denote the probability of finding the particle on the excited electronic state at $x$ in time $t$ and we use $P_{g}(x,t)$ to denote the probability of finding the particle on the ground electronic state at $x$ in time $t$. In our model de-excitation is modelled by transferring the particle from excited electronic state to the ground electronic state.  We assume that the motion on both the potential energy curve is overdamped, which is very common inliterature. Therefore both the probability $P_{e}(x,t)$ and $P_{g}(x,t)$ may be found at $x$ at the time $t$ obeys a modified Smoluchowskii equation.
\begin{eqnarray}
\frac{\partial P_e(x,t)}{\partial t} = {\cal L}_e P_e(x,t) - k_r P_e(x,t) - k_0 S(x) P_g(x,t)  \\ \nonumber
\frac{\partial P_g(x,t)}{\partial t} ={\cal L}_g P_g(x,t) + k_r P_g(x,t) + k_0 S(x) P_e(x,t) \nonumber
\end{eqnarray}
\noindent In the above 
\begin{equation}
{\cal L}_i= A \frac{\partial^2}{\partial x^2}+\frac{\partial}{\partial x} \frac{dV_i(x)}{dx}.
\end{equation}
\noindent In the above equation, $V_i(x)$ is the potential causing the drift of the particle, for electronically excited state $V_{e}(x) = \frac{1}{2} k (x-a)^2+ b$ and for ground electronic state $V_{g}(x) = \frac{1}{2} k (x+a)^2 - b$, $S(x)$ is a position dependent coupling function, $k_0$ is the rate constant of nonradiative decay process from the electronically excited state and $k_r$ is the rate constant of radiative decay process from the electronically excited state. We have assumed $k_r$ to be independent of position. $k_r P_g(x,t)$ term represent the rate of increase of probability of finding a molecules in the ground electronic state. $A$ is the diffusion coefficient, we have assumed it to be same in both the potentials. Before we excite, the molecule is in the ground electronic state, and as the solvent is at the finite temperature T, its distribution over the coordinate $x$ is random. From this the molecule undergoes Franck-Condon excitation to the electronically excited state. Therefore, $x_0$ the initial position of the particle, on the excited state potential energy surface is also random. We take it to be given by the probability density $P^{0}_{e} (x)$. In the following we provide a general procedure for finding the exact solution of Eq. (1) in time domain. Now Eq. (1) can be written as
\begin{eqnarray}
\frac{\partial P_e(x,t)}{\partial t} = A \frac{\partial^2 P_e(x,t)}{\partial x^2}+ k \frac{\partial}{\partial x} x  P_e(x,t) - (k_r+a) P_e(x,t) - k_0 S(x) P_g(x,t)  \\ \nonumber
\frac{\partial P_g(x,t)}{\partial t} = A \frac{\partial^2 P_g(x,t)}{\partial x^2}+ k \frac{\partial}{\partial x} x  P_g(x,t)  + (k_r+a) P_g(x,t) + k_0 S(x) P_e(x,t) \nonumber
\end{eqnarray}
\noindent Now the above two equation can be combined to write a single equation as given below
\begin{eqnarray}
\frac{\partial P(x,t)}{\partial t} = A \frac{\partial^2 P(x,t)}{\partial x^2}+ k \frac{\partial}{\partial x} x  P(x,t) - (k_r+a) P(x,t) - i k_0 S(x) P(x,t).
\end{eqnarray}
\noindent where $P(x,t) = P_e(x,t) - i P_g(x,t)$. therefore if we separate the real part and imaginary part of the above equation, we get those two equations which are written in Eq.(3). In the following we will solve the above equation to find an analytical expression of $P(x,t)$ and real part of $P(x,t)$ corresponds to $P_e(x,t)$ and imaginary part corresponds to $P_g(x,t)$. We find it convenient to start with the above equation without any decay term and the corresponding equation is given by
\begin{equation}
\frac{\partial P(x, t)}{\partial t} = D\frac{\partial^2 P(x,  t)}{\partial x^2} + k  \frac{\partial}{\partial x} \left( x P(x,  t)\right),
\label{eqn:harmonic}
\end{equation}
\noindent For Dirac delta function $P(x,0) = \delta (x)$, the solution of the above equation is known to be,
\begin{equation}
P(x, t) = \frac{e^{-\frac{(x)^2}{4 D \sigma(t)^2}}}{ \sqrt{4 D \pi}\sigma(t)} ,
\label{eqn:p0}
\end{equation}
\noindent where $\sigma(t)^2 = \frac{1}{2k}(1 - e^{- 2k t})$. Rearranging the Eq. \ref{eqn:harmonic} gives,
\begin{equation}
\frac{\partial P(x, t)}{\partial t} - k \frac{\partial}{\partial x} \left( x P(x,  t)\right) = D \frac{\partial^2 P(x,  t)}{\partial x^2}.
\label{eqn:3}
\end{equation}
\noindent Now, we insert the $P(x,t)$ given by Eq. \ref{eqn:p0} to compute the L. H. S. of Eq. \ref{eqn:3},
\begin{equation}
\frac{\partial P(x, t)}{\partial t} -  k\frac{\partial}{\partial x} \left( x P(x,  t)\right) = \frac{-Dk(1-e^{-2kt})+k^2x^2}{D(1-e^{-2kt})^2} P(x,t),
\end{equation}
\noindent while substituting the $P(x,t)$ given by Eq. \ref{eqn:p0} gives the R. H. S. of Eq. \ref{eqn:3} as,
\begin{equation}
D\frac{\partial^2 P(x,  t)}{\partial x^2} = \frac{-Dk(1-e^{-2kt})+k^2x^2}{D(1-e^{-2kt})^2} P(x,t).
\end{equation}
\noindent Now we take only the time derivative of $P(x,t)$, and we get
\begin{equation}
 \frac{\partial P(x, t)}{\partial t}  =  e^{-2kt}\frac{-Dk(1-e^{-2kt})+k^2x^2}{D(1-e^{-2kt})^2} P(x,t),
\end{equation}
\noindent By combining the above three equations, the following relation can be written
\begin{equation}
e^{2k t} \frac{\partial P(x, t)}{\partial t}  =  D\frac{\partial^2 P(x,  t)}{\partial x^2}
\end{equation}
\noindent where $P(x,t)$ is given by Eq. \ref{eqn:harmonic}. Now we can re-write the above equation as
\begin{equation}
\frac{\partial P(x,\tau)}{\partial \tau}  =  D \frac{\partial^2 P(x, \tau)}{\partial x^2}
\label{eqn:flat}
\end{equation}
\noindent  Again $P(x,t)$ is given by Eq. \ref{eqn:harmonic},
\begin{equation}
\frac{\partial }{\partial \tau}= \frac{\partial t }{\partial \tau}
\frac{\partial }{\partial t},
\end{equation}
which implies,
\begin{equation}
\frac{\partial t }{\partial \tau} = e^{2 k t},
\end{equation}
and therefore the variable $\tau$ is derived to be
\begin{equation}
\int_{0}^{\tau} d \tau' = \int_{0}^{t} e^{- 2 k t'} dt'  \implies \tau = \frac{1- e^{- 2k t}}{2k}. 
\end{equation}
\noindent The solution of the flat potential yields the solution for harmonic potential by replacing $\tau$ by $\frac{1-e^{-2kt}}{2k}$. The solution of Eq. \ref{eqn:flat} is given by,
\begin{equation}
P(x, \tau) = \frac{e^{ - \frac{(x)^2}{4 D \tau}}}{ \sqrt{4 \pi D \tau}} ,
\end{equation}
Therefore the solution of Eq. (7) can be derived to be,
\begin{equation}
P(x, t) = \frac{e^{-\frac{(x)^2}{2\frac{D}{k}(1-e^{-2kt})}}}{\sqrt{2\frac{D}{k}\pi (1- e^{- 2k t})}} 
\end{equation}
\noindent Now we will add all the decaying term in Eq.\ref{eqn:harmonic} and assume $S(x) = \delta(x-x_c)$ to get
\begin{equation}
\frac{\partial P(x, t)}{\partial t} = D\frac{\partial^2 P(x,t)}{\partial x^2} +k  \frac{\partial}{\partial x} \left(x P(x,t)\right) - k_r' P\left(x,t\right) - i k_{0}\delta(x-x_c)P\left(x,t\right)
\label{eqn:harmonicdd}
\end{equation}
\noindent with $k_r'= k_r+a$. The above equation may be written as
\begin{equation}
 \frac{\partial P(x,\tau)}{\partial \tau} =D\frac{\partial^2 P(x,\tau)}{\partial x^2}  - k_r' P(x,\tau) - i k_{0} \delta (x - x_c) P(x,\tau),
\end{equation}
Now we solve the above equation for flat potential and in the solution we will perform appropriate replacements to obtain the solution of Eq. \ref{eqn:harmonicdd}. The above equation can be solved using the half-Fourier transformation,
\begin{equation}
\tilde P(x,\omega)= \int^\infty_0 P(x,\tau) e^{i \omega \tau} d\tau.
\end{equation}
Laplace transformation of Eq. (1) yields the following equation
\begin{equation}
-i \omega {\tilde P}(x,\omega)-P(x,0)=D \frac{\partial^2{\tilde P}(x,\omega)}{\partial {x}^2} - k_r' {\tilde P}(x,\omega) - i k_0 \delta(x -{x}_c){\tilde P}(x,\omega).
\end{equation}
\noindent We use $P(x,0) = \delta(x)$,
\begin{equation}
-i \omega {\tilde P}(x,\omega)-\delta(x)= D \frac{\partial^2{\tilde P}(x,\omega)}{\partial {x}^2} - k_r' {\tilde P}(x,\omega) - i k_0 \delta(x -{x}_c){\tilde P}({x}_c,\omega).
\label{eqn:freefourier}
\end{equation}
\noindent Eq. \ref{eqn:freefourier} may be solved using the Green's function method and the solution is given by  
\begin{equation}
{\tilde P}(x,\omega)= \int^\infty_{-\infty} d{x}_{i}{\tilde G}_{0}(x,{x}_i|\omega+i k_r')\delta(x_i) - i k_0 {\tilde P}({x}_c,\omega)\int^{\infty}_{-\infty} dx_{i}{\tilde G}_{0}(x,x_i|\omega + i k_r')\delta(x_i -{x}_c).
\end{equation}
\noindent On simplification
\begin{equation}
{\tilde P}(x,\omega)={\tilde G}_{0}(x,0|\omega + i k_r') - i k_0 {\tilde P}({x}_c,\omega){\tilde G}_{0}(x,x_c|\omega + i k_r').
\end{equation}
\noindent The unknown ${\tilde P}(x_c,\omega)$ can be solved by substituting $x = x_c$ and is obtained to be,
\begin{equation}
{\tilde P}({x}_c,\omega)= \frac{{\tilde G}_{0}(x_c,0|\omega + i k_r')}{1 + i k_0 {\tilde G}_{0}(x_c,x_c|\omega + i k_r')}.
\end{equation}
\noindent Now we derive the distribution in the Fourier domain,
\begin{equation}
{\tilde P}({x},\omega)={\tilde G}_{0}({x},0|\omega + i k_r') - \frac{i k_0 {\tilde G}_{0}(x,x_c|\omega + i k_r') {\tilde G}_{0}(x_c,0|\omega + i k_r')}{{1 + i k_0 {\tilde G}_{0}(x_c,x_c|\omega + i k_r')}}.
\end{equation}
\noindent The Green's function without the Dirac delta sink term is given by,
\begin{equation}
{\tilde G}_{0}(x,x_i|\omega + i k_r') = \frac{e^{- \sqrt{\frac{-i \omega + k_r'}{D}}|x -x_i |}}{2\sqrt{(- i \omega + k_r')D}},
\end{equation}
\noindent which gives the following expression of  $P(x,\omega)$,
\begin{equation}
{\tilde P}(x,\omega )= \frac{e^{- \sqrt{\frac{- i \omega + k_r'}{D}}|x |}}{2\sqrt{( - i \omega + k_r')D}}   -\frac{i k_0 e^{-\sqrt{\frac{ - i \omega + k_r'}{D}}(|x|+|x - x_c|)}}{2\sqrt{(- i \omega + k_r') D}(2 \sqrt{ (- i \omega + k_r') D} + i k_0)}.
\end{equation}
\noindent The inverse Fourier transformation leads to the expression for $P(x,\tau)$ as given by,
\begin{equation}
P(x,\tau) = \frac{e^{-\frac{x^2}{4 D \tau} - k_r' \tau}}{2 \sqrt{\pi D \tau}}-\frac{i k_0}{4D}e^{\frac{- k_0^2}{4 D} \tau 
+\frac{i k_0}{2D}(|x-x_{c}|+|x_c|)} Erfc{\frac{i k_0}{2\sqrt{D}}\sqrt{\tau}+\frac{|x-x_c|+|x_c|}{2\sqrt{D \tau}}}.
\end{equation}
\noindent Now we replace $\tau$ by $\frac{1 - e^{- 2 k t}}{2 k}$ to get the solution of the following equation
\begin{equation}
\frac{\partial P(x, t)}{\partial t} = D\frac{\partial^2 P(x,t)}{\partial x^2} + k  \frac{\partial}{\partial x} \left(x P(x,t)\right) - k_r' P\left(x,t \right) - k_{0}\delta(x-x_c)P\left(x,t \right),
\label{eqn:harmonicddnew}
\end{equation}
\noindent Therefore we get
\begin{eqnarray}
   P(x,t)= \frac{e^{-\frac{x^2}{4 D \frac{1 - e^{- 2 k t}}{2 k}} - k_r' t}}{2 \sqrt{\pi D \frac{1 - e^{- 2 k t}}{2 k}}}-\frac{i k_0}{4D}
   e^{\frac{- k_{0}^2}{4D}\frac{1-e^{-2kt}}{2k}+\frac{i k_0}{2D}(|x-x_c|+|x_c|)}
   Erfc{\frac{i k_0}{2\sqrt{D}}\sqrt{\frac{1-e^{-2kt}}{2k}}+\frac{|x-x_c|+|x_c|}{2\sqrt{D \frac{1-e^{-2kt}}{2k}}}}.
\end{eqnarray}
\noindent The survival probability can be obtained by integrating the distribution over all $x$, {\it i.e.}, 
$Q(t) = \int_{-\infty}^{\infty}P(x,t) dx$, which is given by
\begin{eqnarray}
Q(t)=e^{- k_r' t}[1 + e^{\epsilon_0 x_c + \epsilon_{0}^2 D \frac{1- e^{-2 k t}}{2 k}}
\left(Sign(\frac{x_c} {2 D \frac{1-e^{-2kt}}{2k}}+\epsilon_0)- Erf{\frac{x_c+2\epsilon_0 D t}{2\sqrt{D\frac{1-e^{-2kt}}{2k}}}}\right)\nonumber \\
+2\theta(-x_c-2\epsilon_0 D \frac{1-e^{-2kt}}{2k})e^{-\epsilon_0^2 D \frac{1-e^{-2 k t}}{2 k}}-Erfc{\frac{x_c}{\sqrt{4D \frac{1-e^{-2kt}}{2k}}}}].
\end{eqnarray}
\noindent Again $Q(t) = \int_{-\infty}^{\infty}P_{e}(x,t) dx - i \int_{-\infty}^{\infty}P_{g}(x,t) dx = Q_{e}(t) - i Q_{g}(t)$. Therefore real part of $Q(t)$ gives the expression of $Q_{e}(t)$ and imaginary part of $Q(t)$ gives the expression of $Q_{g}(t)$. In summary we gave exact analytical expression of time domain solution for the problem of electronic relaxation in solution.


\begin{thebibliography}{9}



\bibitem{Rice-Book} S. A. Rice, {\it Diffusion Limited Reactions} (Elsevier, Amsterdam, 1985).

\bibitem{Hynes-Rev} J. T. Hynes,  Ann. Rev. Phys. Chem. {\bf 36}, {\it 573} (1985).   

\bibitem{Wolynes} G. R. Fleming and P. G. Wolynes, Phys. Today {\bf 43(5)}, {\it 36} (1990).  

\bibitem{Sadhukhan} A. Samanta and S. K. Ghosh and H. K. Sadhukhan, Chem. Phys. Lett. {\bf 168}, {\it 410} (1990).  

\bibitem{Sutin} R. A. Marcus and N. Sutin, Biochem. Biophys. Acta {\bf 811}, {\it 265} (1985). 

\bibitem{Sumi} H. Sumi and R. A. Marcus, J. Chem. Phys. {\bf 84}, {\it 4894} (1986).
  

\bibitem{Fleming} B. Bagchi and G. R. Fleming, J. Phys. Chem.  {\bf 94}, {\it 9} (1989).    

\bibitem{Oxtoby} B. Bagchi and G. R. Fleming and D. W. Oxtoby, J. Chem. Phys.  {\bf 78}, {\it 7375} (1983).

\bibitem{Bagchi} B. Bagchi, J. Chem. Phys. {\bf 87}, {\it 5393} (1987).  

\bibitem{Szabo} A. Szabo and G. Lamm and G. H. Weiss, J. Stat. Phys.  {\bf 34}, {\it 225} (1984).

\bibitem{Kls} K. L. Sebastian, Phys. Rev. A  {\it 46}, {\it R1732}  (1992). 

\bibitem{Skg} A. Samanta and S. K. Ghosh, Phys. Rev. E {\it 47}, {\it 4568} (1993).

\bibitem{Hynes} J. T. Hynes, J. Phys. Chem. {\bf 90}, {\it 3701} (1986).

\bibitem{Ani_Arb} Diwaker and A. Chakraborty. Mol. Phys. (in press) (2012).

\bibitem{AniJCP} A. Chakraborty, J. Chem. Phys., 139, 094101 (2013).


 

\end{thebibliography}
\end{document}